\documentclass[prb,aps]{revtex4}

\usepackage{color,graphicx,epsfig}
\begin{document}
%\draft
\title{Relaxation of an Electron System : Conserving Approximation}
\author{G. S. Atwal and N. W. Ashcroft}
\affiliation{Cornell Center for Materials Research, and the Laboratory of
Atomic and Solid State Physics, Cornell University, Ithaca NY 14853-2501}
\date{\today}
\setlength{\topmargin}{0in}
\setlength{\headheight}{0in}
\setlength{\headsep}{0in}

\newcommand{\nid}{\noindent}
\newcommand{\ds}{\displaystyle}
\newcommand{\ha}{\widehat}
\newcommand{\hb}{\hbar}
\newcommand{\la}{\lambda}
\newcommand{\ep}{\epsilon}
\newcommand{\cd}{\cdot}
\newcommand{\de}{\delta}
\newcommand{\na}{\nabla}
\newcommand{\om}{\omega}
\newcommand{\lan}{\langle}
\newcommand{\ran}{\rangle}
\newcommand{\pder}[2]{\frac{\partial{#1}}{\partial {#2}}}
\newcommand{\cder}[1]{\frac{D {#1}}{Dt}}
\newcommand{\pdder}[2]{\frac{{\partial}^2{#1}}{\partial {#2}^2}}
\newcommand{\bearraynn}{\begin{eqnarray}}
\newcommand{\eearraynn}{\end{eqnarray}}
\newcommand{\benn}{\begin{equation}}
\newcommand{\eenn}{\end{equation}}
\newcommand{\eq}[1]{{Eq.~(\ref{#1})}}
\newcommand{\eqs}[2]{{Eqs.~(\ref{#1}--\ref{#2})}}
\newcommand{\lab}{\label}
\newcommand{\frd}[2]{\ds \frac{{#1}}{{#2}}}
\newcommand{\cn}[2]{\chi_{#1}({\bf q},{#2})}
\newcommand{\wit}{\tilde{\om}}
\newcommand{\ppq}{{\bf p}\! + \! \frac{1}{2}{\bf q}}
\newcommand{\pmq}{{\bf p}\! - \! \frac{1}{2}{\bf q}}
\newcommand{\dep}{\lan \ppq | \ha{\de \rho} | \pmq \ran}
\newcommand{\De}{\Delta}
\newcommand{\wmpq}{\frac{\om m}{p\, q}}
\newcommand{\pqwm}{\frac{p\, q}{\om m}}
\newcommand{\pfqwm}{\frac{p_F q}{\om m}}
\newcommand{\wmpfq}{\frac{\om m}{p_F q}}

%\maketitle

\small

\begin{abstract}
The dynamic response of an interacting electron system is determined by an
extension of
the relaxation-time approximation forced to obey local conservation laws for
number, momentum and energy. A consequence of these imposed constraints is
that the local
electron equilibrium distribution must have a space- and time-dependent
chemical
potential, drift velocity and temperature. Both quantum kinetic and
semi-classical
arguments are given, and we calculate and analyze the corresponding
analytical $d$-dimensional
dielectric function. Dynamical correlation, arising from relaxation
effects, is shown to soften the plasmon dispersion of both two- and
three-dimensional systems. Finally, we consider the consequences for a
hydrodynamic theory of a
$d$-dimensional interacting electron gas, and by incorporating the
competition between relaxation and inertial effects we derive generalised
hydrodynamic
equations applicable to arbitrary frequencies.

\end{abstract}

%\begin{multicols}[2]
\maketitle
\section{Introduction}

The dynamic response of correlated and/or damped electron systems has both
received and
stimulated considerable theoretical work and continues to present
challenges. The dynamic
dielectric function (DF) of an undamped quantum electron gas was first
calculated in the
random phase approximation (RPA) by Lindhard \cite{Lindhard}. His approach
assumed an
infinite electron relaxation time, $\tau \to \infty$, and it was suggested
that the
effects of phenomenological damping could be incorporated by assuming a finite
relaxation time. However it is well known that this single relaxation-time
approximation
fails to locally conserve the basic conservation laws for electron number,
momentum and
energy. This results in a number of incorrect experimental predictions such as
alteration of the static DF even though damping is but a dynamic phenomenon.

The first corrective step taken to rectify this situation was carried out by
Mermin
\cite{Mermin} who was able to derive a DF which conserved electron number
during
collisions. This was achieved by using a relaxation-time approximation in
which the
collisions relax the driven electron distribution not to its global equilibrium
distribution, but to a local equilibrium distribution specified by a local
chemical
potential, $\mu({\bf r},t)$. This number-conserving approximation has since
been widely
used to study the effects of scattering in several different systems such
as periodic
crystal potentials \cite{Ash1}, small metallic particles \cite{Ash2} and
interacting
storage ring plasmas \cite{Morawetz}. Nevertheless, the Mermin DF violates
the two
remaining conservation laws and, in a one-component system of electrons,
any mechanism
which induces relaxation of a non-equilibrium electron distribution may not
violate any
of the three laws. Even if, in the presence of external sources, momentum
loss does
occur then energy conservation may not necessarily be affected as in the
case of
non-magnetic static impurities. In this paper we show how satisfaction of all
three conservation laws, and various combinations thereof, may be achieved in
determining the dynamic response of an electron gas. The proposed key to
the solution is
that the local equilibrium distribution must not only exhibit spatial and
temporal
variations in $\mu$ but also in the drift velocity, ${\bf v}$ (to conserve
momentum),
and temperature, $T$ (to conserve energy). This idea  \cite{us} has,
independently, also been recently introduced, and implemented in the context of
generalised quantum liquids
\cite{Moranew}.

We begin by explicitly deriving the full conserving dielectric function
(FCDF) as
appropriate for a one-component, $d$-dimensional, quantum system of
electrons. In this
context the relaxation time approximation can be viewed as a way of
incorporating, at
least approximately, dynamical correlation effects within the response of
an interacting
electron gas. The corrections to the non-collisional DF, arising from the
conserving
(relaxation-time) approximation \cite{Kadanoff}, thus serve the same
purpose as the much
sought-after dynamical local field corrections of the DF of an interacting
electron gas.
Other straightforward applications of the conserving approximation, that
will be
discussed, include to the plasmon dispersion and conductivity of the electron system.

We will also demonstrate that a correct accounting of relaxation effects is of
considerable importance in providing an accurate hydrodynamic description
of an electron
gas. Though hydrodynamic models of an electron gas \cite{Bloch} are popular
by virtue of
their simplicity, there remains a long-standing problem in that they
may predict physically
incorrect plasmon dispersion relationships in both two and three
dimensions. The difficulties can be traced back to the assumption of
adiabacity, valid only at low frequencies, where there is an
overcompensation of relaxation effects. Inertial effects
become important at higher frequencies and thus a stress tensor is
required, in contrast
to specification of the sole scalar pressure term used in traditional
adiabatic hydrodynamics. This is
particularly important in the case of three-dimensional plasmons where the
strong
Coulomb potential is responsible for a large frequency gap at zero
wavevector. In contrast,
relaxation effects from electron-electron scattering compete more
strongly at lower frequencies to maintain local stress
isotropy. It follows that a hydrodynamic model of an electron gas,
generalised to arbitrary
frequency, must successfully incorporate competition between both
effects. Tokatly and Pankratov \cite{Tokatly1} have also recently
succeeded in deriving such a model for a
three-dimensional electron gas based on the Boltzmann transport equation,
and have gone
on to show how such generalised hydrodynamics can be applied to the Landau
theory of
Fermi liquids. Here we present a derivation of the model for a $d$-dimensional electron
gas starting from the microscopic quantum dynamics but with the relaxation-time
approximation forced to conserve all three laws, and by way of example,
demonstrate the manner in
which plasmon dispersions are then correctly deduced.

The plan of this paper is as follows : in Sec. \ref{sec:qu} we introduce a
relaxation-time approximation within a formalism of quantum transport
theory which
allows for local chemical potential, drift velocity and temperature
variations. In Sec.
\ref{sec:die} we develop applications of the conserving relaxation-time
approximation by,
firstly, deriving the new DF and comparing its properties, including
plasmon dispersion,
with previous models, and secondly, considering the conductivity. In Sec.
\ref{sec:semi}
we show how an analogous semi-classical derivation can be made, and in Sec.
\ref{sec:hyd} we derive a hydrodynamic model of an electron gas generalised
to arbitrary
frequencies.

\section{Quantum Kinetics} \lab{sec:qu}

\subsection{Model}

The model we consider, primarily, is a system of charged electrons embedded
in a
neutralizing background (jellium model), whereby electrons are only
scattered by other electrons, and the dynamics of such scattering events are
constrained by all the conservation laws. In any realistic setting
there will also be phonons and impurities giving rise to additional
scattering mechanisms with fewer conservation constraints, and indeed
the loss of
momentum and energy is important in a discussion of the conductivity and
dissipation of
the system under the influence of an external field. If all
possible scattering processes are treated at the same phenomenological
level we need only to discuss in detail the more constrained problem of a
one-component
system (i.e. electron-electron scattering only) to understand how
conservation constraints may be applied in general to the other scattering
mechanisms. The one-component model has the additional virtue of
allowing us to calculate dynamical local field corrections of the DF
arising entirely from electron-electron correlation effects.

The treatment of a $d$-dimensional many-electron system is simplified
by reduction to an underlying single particle description. In practice this
means tracing
out all the degrees of freedom associated with the other remaining
electrons and embedding
the many-body interaction term in the dynamics of a single electron. In
this way, the
dynamics of a single electron is governed by two terms : i) an effective
single particle
Hamiltonian, and ii) a term describing interactions with the other particles.

\subsection{Relaxation-Time Approximation}

Hence, we begin by considering the dynamics of the quantum one-electron
statistical operator
operator $\ha{\rho}$. The fermionic single electron equilibrium statistical
operator is given
by ($\hb=1$)

\benn \ha{\rho}_0 = \frac{1}{e^{(\ha{H}_0- \ha{\bf k}\cd{\bf v} -
\mu)/k_BT}+1}, \eenn

\nid where $\ha{H}_0$ and $\ha{\bf k}$ are the Hamiltonian and the momentum
operator for
an independent and non-interacting electron. Henceforth we shall refer to
$\ha{\rho}_0$
as the global equilibrium statistical operator. The distribution of a
moving gas differs from
that at rest by a Galilean transformation of the drift velocity ${\bf
v}=\lan{\bf
k}\ran/m$, which is assumed to be much smaller than the Fermi velocity.
With a view to
future simplification we set the global equilibrium drift velocity, ${\bf
v}_{eq}$,
equal to zero (i.e. the change from equilibrium and the magnitude of the
drift velocity
is due to an applied field). Note also that in a crystalline lattice the
mass of an
electron, $m$, which enters the calculations is taken to be the effective
mass thereby
incorporating, at least approximately, band structure. For any eigenstate
$|i \ran$ of
$\ha{H}_0$,

\benn \ha{\rho}_0|i\ran=\frac{1}{e^{({\ep_i}-{\bf k}_i\cd{\bf
v}-\mu)/k_BT}+1}|i\ran=f_i|i\ran, \eenn

\nid where $f_i$ is the Fermi-Dirac occupation function for that state.
Upon application
of an external perturbation, $\ha{U}_{\rm ext}$, and ensuing internal response,
$\ha{U}_{\rm int}$, the dynamics of the total single particle statistical
operator, $\ha{\rho}$, is
governed by the quantum Liouville equation

\benn
\pder{\ha{\rho}}{t}+i[\ha{H},\ha{\rho}]=\left(\pder{\rho}{t}\right)_{\rm
coll},
\lab{Liou} \eenn

\nid where $\ha{H}$ is now the effective Hamiltonian, i.e.

\bearraynn
\ha{H}&=&\ha{H}_0+\ha{U}, \\
\ha{U}&=&\ha{U}_{\rm ext}+\ha{U}_{\rm int}, \nonumber \eearraynn

\nid and the term on the right of \eq{Liou} is the collision term. Note
that $\ha{H}$ is
not the full Hamiltonian since it represents mean-field (Hartree-like) dynamics
containing only one-body operators, whereas the two-body terms are subsumed
in the
collision term. The standard relaxation-time approximation is now invoked,
namely

\benn \left(\pder{\rho}{t}\right)_{\rm
coll}=-\frac{\ha{\rho}-\ha{\rho}_{\rm loc}}{\tau}. \lab{collterm}
\eenn

The usual interpretation of the approximation is as
follows : the electron ensemble evolves in
time as determined by the left-hand side of the Liouville equation,
\eq{Liou}, but in a time
interval $\de t$ a fraction, $\de t/ \tau$, of them collide and are
redistributed
according to a local equilibrium statistical operator, $\ha{\rho}_{\rm
loc}$. The energy
dependence of $\tau$ can be ignored if we assume that the electron levels
involved in the
collision processes lie close to the Fermi energy.

However, the familiar approximation of setting $\ha{\rho}_{\rm loc}$ equal to
$\ha{\rho}_0$ leads to a violation of conservation laws.
To determine the correct (dynamical) local equilibrium state we observe
from classical
thermodynamics that any state of an ensemble of particles can be specified
by five
parameters (in three dimensions), two coming from the thermodynamic
variables $\mu$ and $T$,
and three from ${\bf v}$. These parameters are well defined
quantities (constants) in a global equilibrium state, as specified by
$\ha{\rho}_0$. If we
now allow these thermodynamic parameters to become functions of space
and time, such that there is no entropy production, then the entire
electron gas is said to be
near-equilibrium and locally the electron distribution is given by
the form of the global equilibrium distribution, $\ha{\rho}_0$, but now
with local
parameters. These parameters must be constrained, and hence determined by,
conservation
laws for average particle number, momentum and energy. In other words, by allowing
local
fluctuations of $\mu$, ${\bf v}$ and $T$ we gain the necessary degrees of
freedom to
satisfy the conservation laws. The thermodynamics of the local equilibrium
state and the
physical meaning of thermodynamic quantities in a near-equilibrium gas is
further
elaborated in Appendix \ref{app:localtherm}.

As appropriate for linear response theory we treat the varaiations of the
local
equilibrium state as small, and thus the local Hamiltonian specifiying the
local
distribution, can be expanded as

\benn \ha{\rho}_{\rm loc}(\ha{H}_{\rm loc}) = \ha{\rho}_0(\ha{H}_0) + \la
\ha{\rho}_1 +
O(\la^{2}), \eenn \benn \ha{H}_{\rm loc}= \ha{H}_0 + \la \left[- \de \mu -
\ha{\bf k} \cd \de {\bf v} - \frac{(\ha{H}_0 -\mu)}{T} \de T \right]
+O(\la^{2}), \eenn

\nid where $\la$ is linked to the order of the expansion. By
definition, the local equilibrium density operator must commute with
the local Hamiltonian and thus we obtain the following matrix element to
first order
(after setting $\la=1$),

\benn \lan{\bf k}|\ha{\rho}_{\rm loc}|{\bf k'}\ran=f_{\bf k} \de_{{\bf kk}'} -
\left( \frac{f_{\bf k} - f_{{\bf k}'}}{\ep_k -
\ep_{k'}} \right) \left[ \de \mu({\bf k}-{\bf k'}) + \frac{{\bf k} +{\bf
k}'}{2} \cd \de {\bf v}({\bf
k}-{\bf k'}) +\frac{\ep_{\bf k}+\ep_{\bf k'}-2\mu}{2T} \de T({\bf k}-{\bf
k'}) \right].
\eenn

\nid Note that we have implicitly assumed the electron state distribution
to be isotropic and
homogeneous in ${\bf k}-$space. To determine the response we assume that
the perturbing
potential, and hence other dynamic quantities, varies as $e^{i({\bf q}\cd
{\bf r} -\om
t)}$. We restrict our analysis to a longitudinal response such that
$\na U_{\rm ext} \times {\bf q}=0$. The dynamics of the first order density
matrix, $\ha{\de \rho}$, is then determined by \eq{Liou} and \eq{collterm}
giving

\bearraynn \om \lan{\bf k}|\ha{\de \rho}|{\bf k'}\ran&-&(\ep_{\bf
k}-\ep_{\bf k'}) \lan{\bf k}|\ha{\de
\rho}|{\bf k'}\ran+(f_{\bf k'}-f_{\bf k})U({\bf k}-{\bf k}')= \\
\nonumber &&-\frac{i}{\tau} \left\{ \lan{\bf k}|\ha{\de \rho}|{\bf
k'}\ran+\frac{f_{\bf k}-f_{\bf k'}}{\ep_{\bf k}-\ep_{\bf k'}}\left[\de \mu({\bf k}-{\bf
k'}) + \frac{{\bf k}
+{\bf k}'}{2} \cd \de {\bf v}({\bf k}-{\bf k'}) +\frac{\ep_{\bf k}+\ep_{\bf
k'}-2\mu}{2T} \de T({\bf k}-{\bf k'}) \right] \right\}. \lab{lineqmo}
\eearraynn

\nid It is convenient to change variables, ${\bf k}\to{\bf
p}+\frac{1}{2}{\bf q}$ and ${\bf
k'}\to{\bf p}-\frac{1}{2}{\bf q}$, thereby making apparent the disturbance
wavevector,
${\bf q}$, i.e.

\bearraynn (\om-\De\ep)&& \dep +\De f U({\bf q})=
\\ \nonumber &&-\frac{i}{\tau} \left\{ \dep
+\frac{\De f}{\De \ep}\left[\de \mu({\bf q}) + {\bf p} \cd \de {\bf v}({\bf q})
+\left( \frac{\De \ep}{2}+\ep_{\pmq}-\mu \right) \ds\frac{\de T({\bf
q})}{T} \right]
\right\}, \lab{aeqmo} \eearraynn

\nid where $\De \ep\ \equiv \ep_{\ppq}-\ep_{\pmq}$ and $\De f\equiv
f_{\ppq}-f_{\pmq}$
are the single-particle changes in the energy and distribution function
respectively.

\subsection{Conservation Laws}
\lab{sec:cons} The expectation values for the conserved quantities (number,
momentum and
energy), in a particular distribution, are obtained by taking the
appropriate moments of the distribution in momentum-space. In general,
momentum and energy will not be fully conserved in the presence of
other degrees of freedom, e.g. phonons and magnetic impurities. In fact,
loss of
momentum is necessary to render a finite conductivity and this will be
discussed later in Sec. \ref{sec:conduct}. Here we will concentrate
on a full-conserving model and to ensure that all these quantities are
collision invariants we must obtain the same expectation values before and
after each
collision. The formal representation of this statement is

\benn \frac{2}{(2 \pi)^d}\int d^d \!p \, \lan{\bf p} + \frac{1}{2}{\bf q}|
\ha{\rho}-\ha{\rho}_{\rm loc} |{\bf p} - \frac{1}{2}{\bf q}\ran \left(
\begin{array}{c}
1 \\
\bf p \\
\ep_p
\end{array} \right) = \left(
\begin{array}{c}
0 \\
0 \\
0
\end{array}
\right). \lab{conslaw} \eenn

\nid Note that the quantities being conserved by \eq{conslaw} are
statistically-averaged, i.e. it is the expectation values of the
density, momentum-density and energy-density which are being
conserved. Of course, angular momentum is also conserved, but in a
non-rotating system of point particles we will not obtain an
additional constraint from this conservation law. This can be seen
from the fact that the angular momentum operator is merely derived
from a bilinear combination of the position and momentum operators, and
thus conservation of angular momentum is automatically enforced.

It follows that the taking of moments of either side
of the linearised equation of motion, \eq{lineqmo}, must necessarily give
zero.  We now
proceed to describe the dynamical response of an interacting electron
gas where the free part has an energy dispersion given by i.e. $\ha{H}_0|{\bf
p} \ran=\ep_{\bf p} |{\bf p} \ran= p^2/2m |{\bf p} \ran$. Enforcement of the
conservation constraints on the right-hand side of \eq{lineqmo} gives

\benn \vec{x}={\mathcal{M}} \vec{f}, \lab{small} \eenn

\nid where $\vec{x}$ is a column vector of the small changes in number
density, momentum
density and energy density, i.e.
\newcommand{\En}{{\mathcal{E}}}

\benn \vec{x}=\left(
\begin{array}{c}
\de n \\
\de {\bf j} \\
\de \En
\end{array} \right)
\equiv \left[ \frac{2}{(2 \pi)^d} \int d^d\!p \lan \ppq | \ha{\de \rho} |
\pmq \ran
\left(
\begin{array}{l}
1 \\
{\bf p} \\
{\ep_p}
\end{array}\right)
\right]. \lab{x} \eenn

\nid Further, \benn \mathcal{M}=\left(
\begin{array}{ccc}
  -B_0 & 0 & (B_0\mu - B_2/2m - B_0q^2/8m)/T \\
  0 & n_0 & 0 \\
  -B_2/2m & 0 & (B_2\mu/2m -B_4/4m^2 -B_2q^2/16m^2)/T
\end{array}
\right), \eenn and \benn \vec{f}=\left(
\begin{array}{c}
  \de \mu \\
  \de {\bf v} \\
  \de T
\end{array}
\right). \lab{mat}
\eenn

For brevity we have also introduced a function, $B_n$, the $n^{{\rm th}}$
momentum
moment of the integrand of the static Lindhard (free-electron)
polarizability function,
namely \benn B_n({\bf q})= \frac{2}{(2 \pi)^d} \int d^d\!p \, |{\bf p}|^n
\frac{\De
f}{\De \ep}, \eenn \nid It will also be useful, in the ensuing analysis, to
introduce
the following related dynamic functions, namely \benn C_n({\bf q},\om)=
\frac{2}{(2
\pi)^d} \int d^d\! p \, |{\bf p}|^n \frac{\De f}{\wit - \De \ep} \eenn \nid
and \benn
D_n({\bf q},\om)=\frac{i C_n - \om \tau B_n}{\wit \tau}, \eenn \nid where
$\wit=\om +
i/\tau$. Analytic expressions for moments of the dynamic $d$-dimensional
Lindhard
polarizability function in the limit $|{\bf q}| \ll k_F$ are evaluated in
Appendix
\ref{app:int}.

To solve for $\vec{f}$ we must obtain a second set of equations, obeyed by
$\vec{x}$,
\eq{x}, by taking the relevant moments of \eq{lineqmo}. We can proceed in
either
of two related ways. The first, is to solve for $\lan \ppq |\ha{\de \rho}|
\pmq \ran$ in \eq{lineqmo} and then take moments, but this results in
rather long and cumbersome intermediate
expressions. We outline such a generalized formalism in the Appendix
\ref{app:lon}. Second, as will be
seen below, we can partially ease the task by noting that the eventual
outcome of all
these calculations is to obtain a self-consistent expression for $\de n$.
In view of
this we need only be concerned with the task of finding alternative
expressions for the
last two components of $\vec{x}$. By imposing conservation of number on the
left-hand
side of \eq{lineqmo} we obtain an equation of continuity, from which we
have \benn \de
{\bf j}=\de n \frac{m \om}{q^2} {\bf q}. \eenn

\nid However there exists no similar strategy for obtaining a simple
expression for $\de \En$ as a function
of $\de n$ and we must resort to the more formal first method, as outlined in
Appendix \ref{app:lon}. This leads to

\benn \de \En=-U({\bf q}) \frac{C_2}{2m}+\frac{B_2+C_2}{2mi\wit \tau}\de \mu +
\frac{C_2}{2mq^2 i \tau}\de{\bf k}\cd{\bf
q}+\left[\frac{B_4+C_2}{4m^2T}+ \left(\frac{q^2}{8m}-\mu
\right)\frac{B_2+C_2}{2mT}\right] \de T.
\eenn

After eliminating the elements of $\vec{x}$ the local thermodynamic variations are then found to be

\benn
 \de \mu({\bf q},\om)=\frac{\left(\ds B_0 2 m \mu - B_2 -
 \frac{B_0 q^2}{4}\right) \left( \ds \frac{i\om C_2}{\tau q^2 n_0} \de n
+UC_2  \right) + \de n \left(D_2
 2m \mu-\ds \frac{D_2 q^2}{4} - D_4 \right)}{D_4 B_0-D_2B_2},
 \lab{demu} \eenn
 \benn
 \de {\bf v}({\bf q},\om)= \de n\frac{\om}{n_0 q^2}{\bf q}, \lab{dev} \eenn and
 \benn \de T({\bf q},\om)= 2mT \frac{\de n \left(D_2 + \ds \frac{i \om B_0
C_2}{q^2 n_0} \right)+
U C_2 B_0}{D_4B_0-D_2B_2}. \lab{det} \eenn

\nid These are among the primary results of this paper; they have immediate
application in the determination of the dynamic DF for a relaxing (correlated)
electron system.

\section{Applications} \lab{sec:die}

\subsection{Dielectric Function} \lab{sec:subdie}

By substituting the expressions for the thermodynamic fluctuations,
\eq{demu}, \eq{dev} and \eq{det}, in the linearized
transport equation, \eq{lineqmo}, we can solve for the density
perturbation, $\de n$. To obtain a self-consistent expression for the
response we specify a Hartree potential for the internal repsonse,
i.e. $U_{\text{int}}({\bf q})=-\de n ({\bf q}) V ({\bf q})$ where
$V({\bf q})$ is the Fourier-transformed Coulomb potential,

\bearraynn V({\bf q}) &=&
2^{d-1}\pi^{(d-1)/2}\Gamma[\frac{1}{2}(d-1)]e^2q^{1-d}, \\
V({\bf q}) &=& \left\{
\begin{array}{c}
4 \pi e^2/q^2, \hspace {1cm} \hbox{(d=3),} \nonumber \\
2 \pi e^2/q, \hspace {1cm}\hbox{(d=2).} \nonumber
\end{array}
\right.
\eearraynn

\nid The longitudinal dielectric function follows immediately
from linear response theory, namely

\benn
\ep({\bf q},\om)=1+V({\bf q}) \frac{\de n({\bf
q},\om)}{U({\bf q})}.
\eenn

\nid The expression for the fully conserving dielectric function (FCDF) is
then found to be

\benn \ep^c({\bf q},\om)=1+V({\bf q})\frac{C_0+E}{1+F}, \lab{gendie} \eenn

\nid where \benn E=\ds \left( \frac{C_2}{\wit i \tau} \right) \ds
\frac{C_2B_0-C_0B_2}{D_4B_0-D_2B_2}
\lab{conE} \eenn and \benn F=\ds \frac{i}{\wit \tau} \left[ \ds
\frac{D_2C_2-D_4C_0+\ds
\frac{i\om C_2}{\tau q^2 n_0}(C_2B_0-C_0B_2)}{D_4B_0-D_2B_2} - 1 \right] +
\ds \frac{i
\om C_0}{\tau q^2 n_0} \lab{conF} \eenn

\nid are the conserving damping corrections. Though cumbersome, some important
properties of the degenerate electron gas are fairly easy to deduce from
\eq{gendie},
\eq{conE} and \eq{conF} with the aid of analytical expressions for certain
integrals
(Appendix \ref{app:int}). First, by taking the limit $\tau \rightarrow
\infty$ in the
FCDF we reassuringly recover the celebrated Lindhard dielectric function in
the absence
of damping, i.e. \benn \ep^0({\bf q},\om)=1+V({\bf q}) \frac{2}{(2 \pi)^d}\int
d^dp\frac{f_{{\bf p} + \frac{1}{2}{\bf q}} -f_{{\bf p} - \frac{1}{2}{\bf
q}}}{\om - {\bf
p} \cd {\bf q}/m +i0^{+}}. \lab{Lind} \eenn

\nid Second, by taking the static limit $\om \rightarrow 0$ we
find \benn \ep^c({\bf q},0)=\ep^0({\bf q},0), \eenn as expected since
relaxation
processes ought to have no bearing on the static properties of the electron
gas. Hence
we also obtain the correct Thomas-Fermi screening result at low {\bf q},
i.e. \benn
q_{\text{TF}}=\left[\frac{d2^{d-2}{\pi}^{(d-1)/2}\Gamma{\bf (} (d-1)/2 {\bf )}
n_0e^2}{E_F}\right]^{1/(d-1)}, \eenn where
$E_F=m^{-1}2^{1-2/d}\pi[\Gamma(1+d/2)]^{2/d}n^{2/d}$ is the $d$-dimensional
Fermi energy. Third, in the long wavelength limit, ${\bf q} \rightarrow 0$,
we retrieve
the same form for $\ep^c$ as that obtained in the absence of damping,
namely \benn
\lim_{{\bf q} \to 0} \ep^c({\bf q},\om)=1-\frd{\om_p^2}{\om^2}, \eenn where
\benn
\om_p^2=\frac{n_0q^2}{m}V({\bf q}), \lab{wp} \eenn

\nid as required by the sum rules for an electron liquid \cite{Pines}.

\begin{figure}
\center{\includegraphics[width=7.5cm]{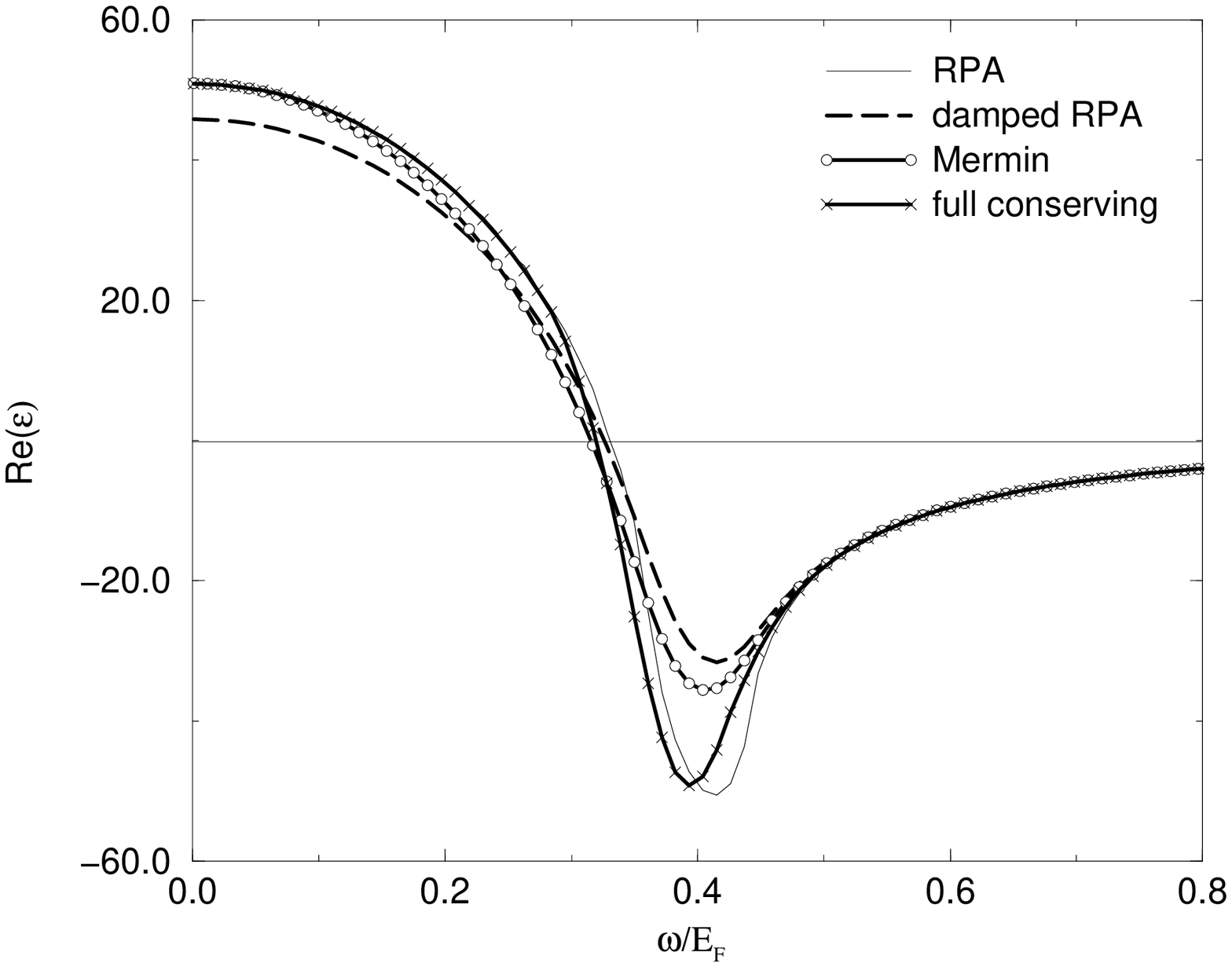}}
\caption{Plot of ${\mathrm Re}(\epsilon)$ against $\omega /E_F$ for a
three-dimensional system. The parameters are
$r_s=3$, $q=(0.2)k_F^{(3d)}$, $\tau=16/E_F^{(3d)}$ and $T=0.01 E_F/k_B$.}
\label{fig1}
\end{figure}

\begin{figure}
\center{\includegraphics[width=7.5cm]{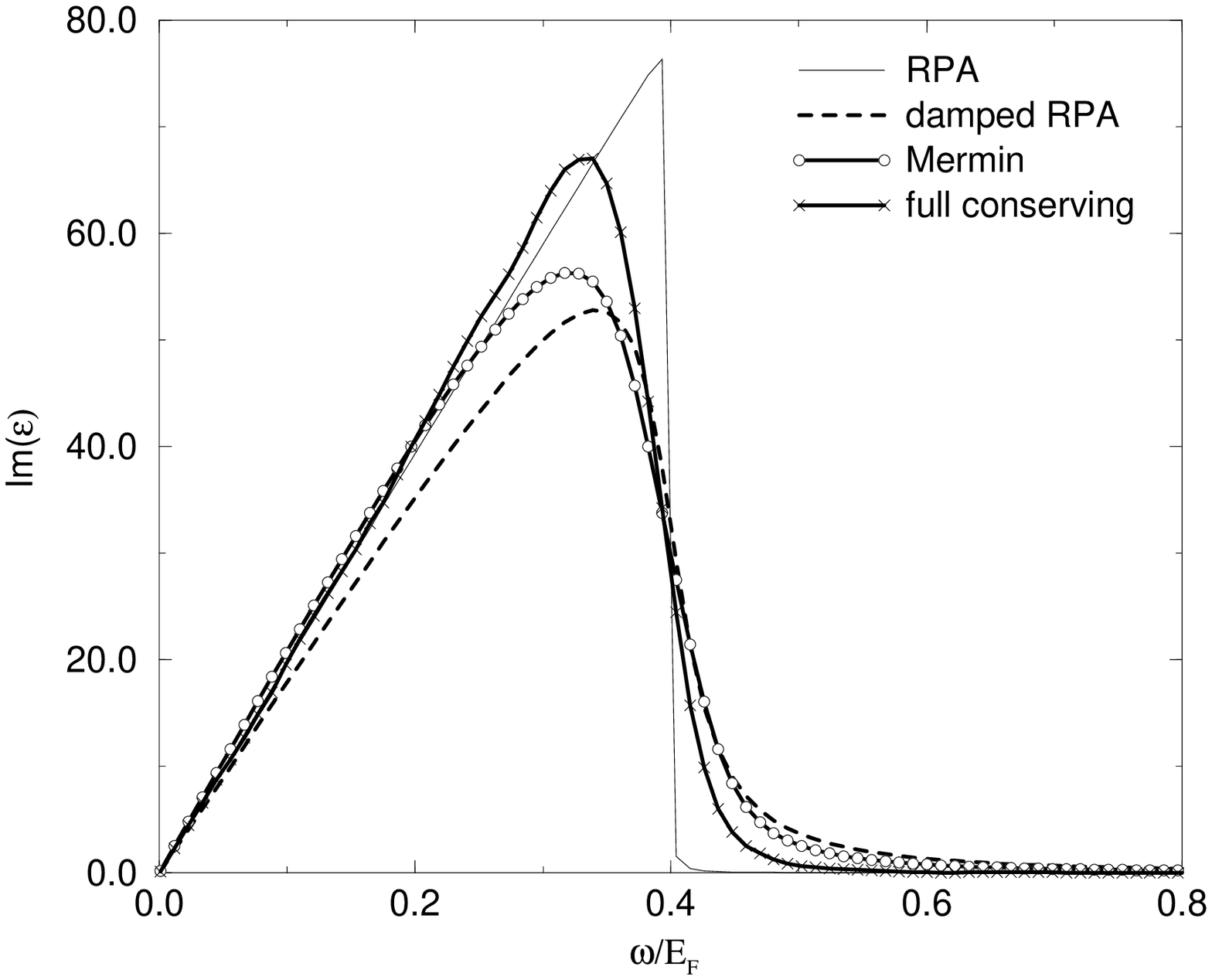}}
\caption{Plot of ${\mathrm Im}(\epsilon)$ against $\omega /E_F$ for a
three-dimensional system. The parameters are the same as in Fig. \ref{fig1}.}
\label{fig2}
\end{figure}

\begin{figure}
\center{\includegraphics[width=7.5cm]{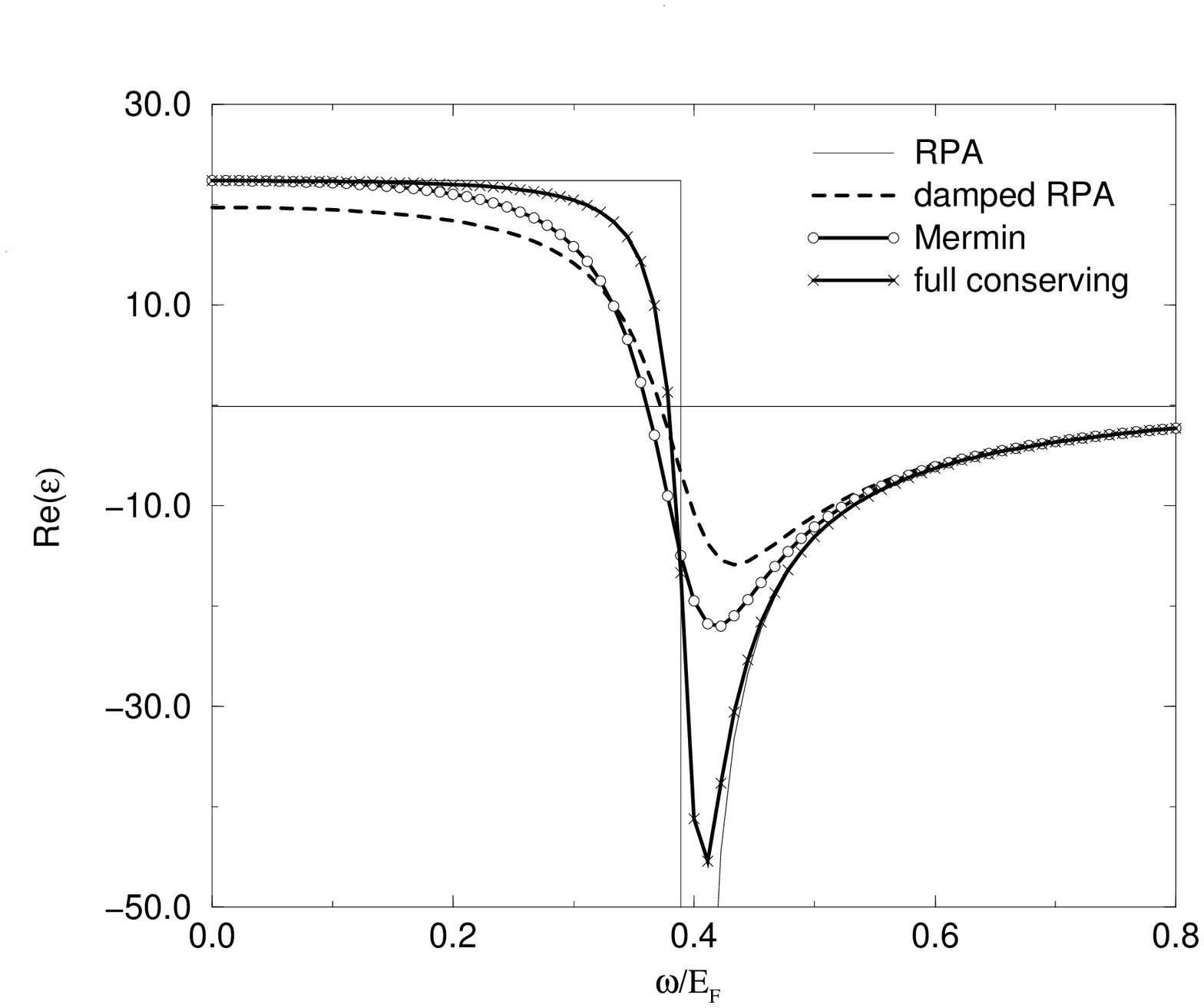}}
\caption{Plot of ${\mathrm
Re}(\epsilon)$ against $\omega /E_F$ for a two-dimensional system. The
parameters are
$r_s=3$, $q=(0.2)k_F^{(2d)}$, $\tau=10/E_F^{(2d)}$ and $T=0.01 E_F/k_B$.}
\label{fig3}
\end{figure}

\begin{figure}
\center{\includegraphics[width=7.5cm]{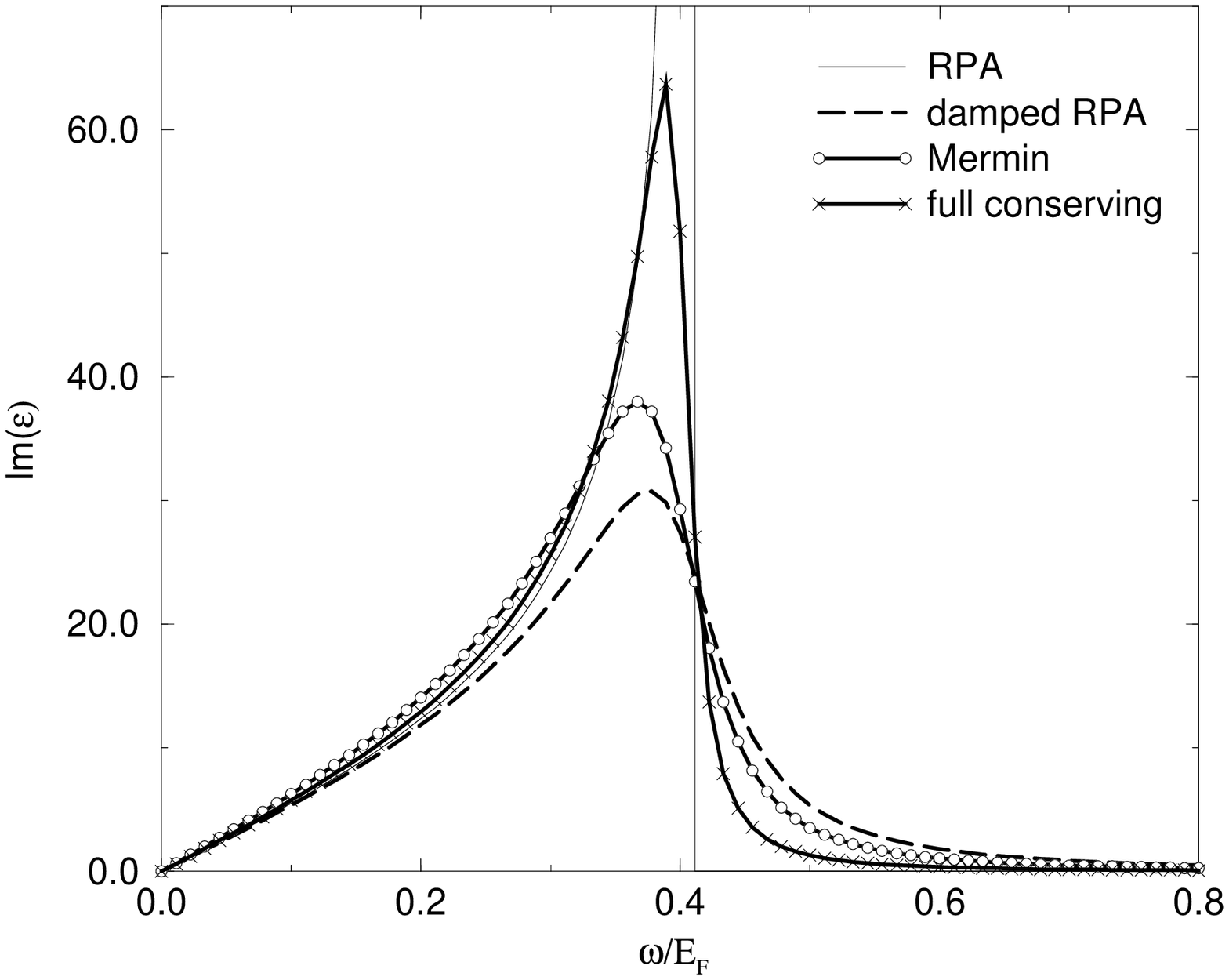}}
\caption{Plot of ${\mathrm Im}(\epsilon)$ against $\omega /E_F$ for a
two-dimensional system. The parameters are the same as in
Fig. \ref{fig3}.}
\label{fig4}
\end{figure}

Using the general form of the dielectric function as given by \eq{gendie}
we may
immediately make comparisons with other prominent models proposed in the
literature :
(a) the undamped Lindhard DF (RPA) corresponding to the choices $E=0$,
$F=0$ and $\tau
\rightarrow \infty$, (b) the damped Lindhard DF: $E=0$, $F=0$ and finite
$\tau$, (c) the
Mermin DF : $E=0$, $F=[i/(\wit \tau)][1+C_0/B_0]$ and finite $\tau$, and
finally (d) the
FCDF given by \eq{conE}, \eq{conF} and finite $\tau$. The real and
imaginary parts of
the various dielectric functions are shown for comparison in three
dimensions (Fig.\ \ref{fig1} and Fig. \ref{fig2}) and two dimensions
(Fig. \ref{fig3} and Fig.\ \ref{fig4}). It is not surprising that in
all cases the behavior of the FCDF most closely
resembles the RPA as it is only in these models that all
the conservation laws are enforced.

\subsection{Plasmon Dispersion}

A simple application of any proposed DF is to determine the dispersion
relationship, $\om(q)$, of
bulk longitudinal plasmons by imposing the condition : $\ep(q,\om)=0$. Plasmon
dispersion data can be obtained by scattering experiments involving X-rays
or electrons.
The measured quantity in these experiments is the energy loss of the
scatterer which is
directly related to the imaginary part of the DF. For the FCDF in the long
wavelength
limit it can be shown that

\benn \lim_{q \to 0} \mathrm{Im}(\ep^{-1})=\frac{\pi}{2}\om_p
\de(\om-\om_p). \eenn

\nid Hence plasmons are the only energy loss at $q=0$ and we have exact
fulfillment of the
long wavelength "perfect screening" sum rule \cite{Pines}:

\benn \lim_{q\to 0} \int_0^\infty \frac{1}{\om} \mathrm{Im}
\ep_l^{-1}(q,\om) d\om =
\frac{\pi}{2}. \eenn

\begin{figure}
\center{\includegraphics[width=7.5cm]{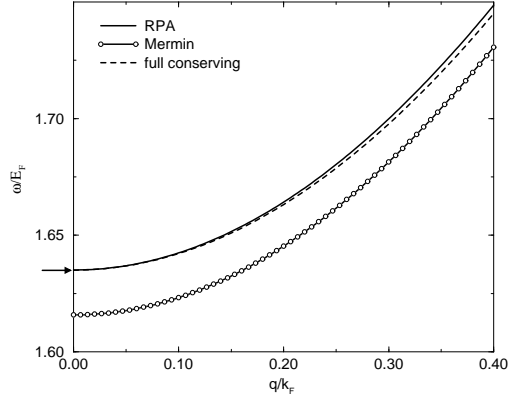}}
\caption{Three-dimensional
plasmon dispersion. The parameters are $r_s=3$, $\tau=3/E_F^{(3d)}$ and $T=0.01
E_F/k_B$. The arrow indicates the plasmon frequency, $\om_p^2=4 \pi e^2
n_0/m$.}
\label{fig5}
\end{figure}

We analyze the solutions for plasmon dispersions arising from the various
dielectric
models as listed in the previous section. For three dimensions the
differing models are
compared in Fig. \ref{fig5}. As expected, the softening of the plasmon
frequency from both the FCDF (at finite $q$) and the Mermin DF
increases with $r_s$ and $1/\tau$. The salient differences between the
models are exhibited in the limit
$q\to 0$. The FCDF gives rise to the same plasmon energy gap, $\om_p$, as
the undamped
Lindhard DF, as required from the sum rules for an electron liquid
\cite{Pines}. In Ref.\ \onlinecite{Morawetz} the Mermin DF was apparently
deduced to obey the perfect screening sum
rule. However this claim can only be true in the limit $\om \to
\infty$ (or $\tau \to \infty$) and it must fail
at finite $\om$ as the 3D plasmon dispersion indicates. In fact, after
a little algebra, it is possible to show that the Mermin DF gives rise to a
Lorentzian spread
in the energy loss,

\benn \lim_{q \to 0} \mathrm{Im} (\ep_{\text{Mermin}}^{-1})=\frac{\om_p^2 \om
/\tau}{(\om^2-\om_p^2)^2+(\om/\tau)^2}. \eenn

\begin{figure}
\center{\includegraphics[width=7.5cm]{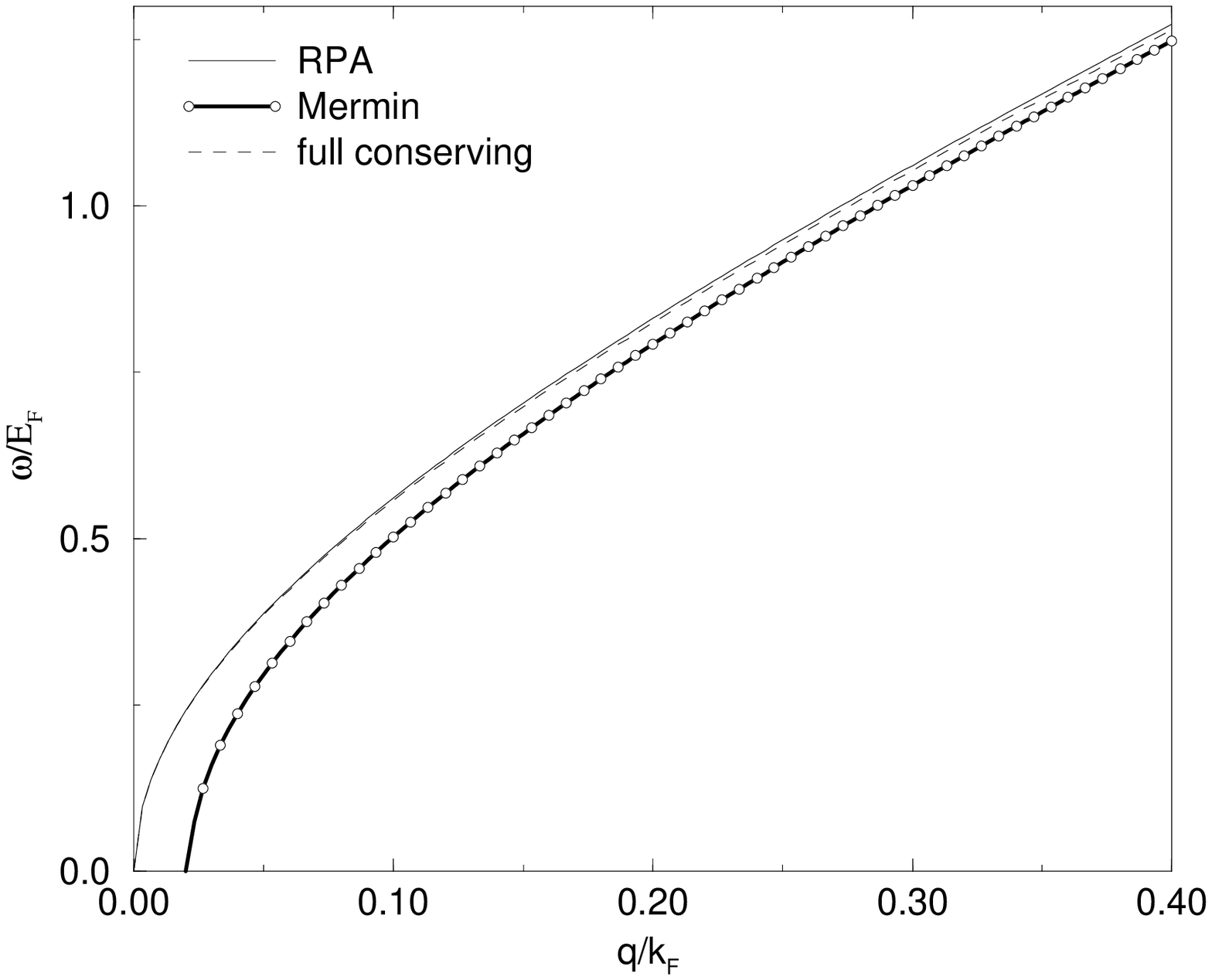}}
\caption{Two-dimensional
plasmon dispersion. The parameters are $r_s=3$, $\tau=5/E_F^{(2d)}$ and $T=0.01
E_F/k_B$.}
\label{fig6}
\end{figure}

The plasmon dispersion for two-dimensional systems is plotted in
Fig.\ \ref{fig6} where it should
be noted that $\tau$ is assumed to be large enough such that
localisation effects are negligible.
As in the three-dimensional case, the plasma excitations from the Mermin DF
and the FCDF
are slightly softened with respect to the plasmon from
RPA. Interestingly, the Mermin DF predicts a wavevector cuttoff below
which a plasma mode cannot exist. This cutoff, which increases with $r_s$ and
$1/\tau$, has also been previously noted \cite{Giuliani} via the
summation of ladder diagrams in the disorderd polarazibility function. The
$q \to 0$ limit of
the FCDF matches that of the RPA DF in agreement with the
aforementioned sum rule. For comparable values of $r_s$ and $\tau$ the
FCDF shows a greater deviation away from the RPA DF in two dimensions
than in three. Clearly, scattering processes
are less important for the three-dimensional plasmon because of the
large frequency gap.

The damping of the two-dimensional plasmon, either from the presence
of external impurities (Mermin DF) or
electron-electron scattering (FCDF), occurs before the onset of Landau
damping (not
shown in figure) and thus may be amenable to experimental verification. In
fact it has
already been appreciated that experimental studies of two-dimensional
electron systems are an
important test of theoretical treatments of many-body effects (beyond RPA)
which have
been shown to be more important in two dimensions than in three
\cite{Giuliani,Jonson}.

\subsection{Conductivity}
\lab{sec:conduct}
In a one-component system, and in an approximation where momentum is
rigorously conserved, we expect a steady state D.C. current to
increase indefinitely upon application of a
constant external field. It can easily be shown that the
conductivity, $\sigma$, as determined by the FCDF, is indeed infinite for the
full-conserving model. To accommodate loss of momentum, thus
rendering a finite conductivity, we return to the equation of motion

\benn \lan \ppq | \ha{\de \rho} | \pmq \ran (\om -{\bf p} \cd {\bf q}/m)
+ \Delta f U({\bf q})=-\frac{i}{\tau} \lan \ppq |
\ha{\rho}-\ha{\rho}_{\text{loc}} | \pmq \ran, \eenn

\nid and impose static ($\om \to 0$) and homogeneous (${\bf q} \to 0$)
conditions, giving (in real space),

\benn \de {\bf j}=\de {\bf j}_{\text{loc}}-n_0 \tau \na U, \eenn

\nid where \benn \de {\bf j}_{\text{loc}}=\frac{2}{(2 \pi)^d} \int
d^d\! p \, e^{i({\bf q} \cd {\bf r} - \om \tau)} \lan \ppq |
\ha{\rho}_{\text{loc}}- \ha{\rho}_0 | \pmq \ran {\bf p}. \eenn

\nid If a fraction, $\lambda$, of the momentum density is conserved
 during collisions, i.e.

\benn \de {\bf j}_{\text{loc}}=\lambda \de {\bf j}, \eenn

\nid then, using $e^2 \de {\bf j}/m=-\sigma \na U$, we
obtain the familiar Drude conductivity

\benn \sigma=n_0 e^2 \tau_{\text{m}}/m, \eenn

\nid where we have defined an effective momentum-relaxation time, \benn
\tau_{\text{m}}=\tau/(1-\lambda). \eenn

\nid As expected, the
conductivity diverges in the conserving limit,
$\lambda \to 1$. In the presence of external scatterers (e.g. fixed
impurities or Umklapp scattering), the other
limit, $\lambda \to 0$, corresponds to
isotropic scattering whereby the net momentum density
is totally destroyed during collisions, and intermediate values of
$\lambda$ reflect the degree of anisotropic scattering.

We may also include a second contribution to the collision integral
stemming from additional scattering processes. These may include
inelastic processes which we have up until now ignored. In other
words, the relevant conductivity relaxation time, $\tau_{\text{cond}}$,
may contain a
contribution from both $\tau$ and an often larger energy relaxation
time \cite{Mahan}, $\tau'$. To demonstrate this we assume, but without
any strict justification, a (decoupled) relaxation time approximation
for the inelastic contribution to the collision integral,

\benn \left(\pder{\rho}{t}\right)_{\rm
coll}=-\frac{(\ha{\rho}-\ha{\rho}_{\text{loc}})}{\tau}
-\frac{(\ha{\rho}-\ha{\rho}_{\text{loc}}')}{\tau'}, \eenn

\nid where $\rho_{\text{loc}}'$ describes a state of local equilibrium
where all
energy and momentum from the motion of particles has been lost. The
expression for
the conductivity then becomes,

\benn \sigma=n_0 e^2 \tau_{\text{cond}}/m, \eenn

\nid where $\tau_{\rm
cond}^{-1}=\tau^{-1}_{\text{m}}+\tau'^{-1}_{\text{m}'}$ (i.e. Matthiessen's
rule).

\section{Semi-Classical Limit} \lab{sec:semi}

We will now derive the dielectric function in a fully classical formalism
except that at
the end of the calculations we will still be free to employ a distribution
appropriate for either
a classical or a fermionic gas. The emerging result for the degenerate
fermionic electron gas, whose equilibrium distribution is governed by
quantum statistics, will be what we call the semi-classical limit.

The central dynamic quantity in classical transport theory is the phase
space electron
distribution $f({\bf p},{\bf r},t)$ governed by the Boltzmann
equation. In the presence of an electric field, ${\bf E}$, the equation of
motion in the
relaxation time approximation is given by

\benn \pder{f}{t}+ \frac{\bf p}{m} \cd \pder{f}{\bf r} + e {\bf E} \cd
\pder{f}{\bf
p}=-\frac{f-f_{\rm loc}}{\tau}. \lab{boltz} \eenn

Magnetic effects are of second order and so disappear when the
equation of motion is linearized.
The total electric force, $e{\bf E}=-\na U$, has a contribution from an
external source
and an internal contribution from the induced charge density perturbation,
as discussed
in the quantum case. To determine the time development we now assume that
the distribution
function is slightly perturbed from equilibrium, and write

\benn f({\bf p},{\bf r},t)=f_0 ({\bf p})+\de f({\bf p},{\bf r},t). \eenn

\nid As in the quantum case we work within the grand canonical ensemble
which serves to define the global
macroscopic temperature and chemical potential. To satisfy the local
conservation laws,
the local equilibrium distribution must be different from the global
equilibrium distribution,
$f_0$, and to first order is given by

\benn f_{\rm loc}=f_0+\pder{f_0}{\mu}\de \mu + \pder{f_0}{\bf v}\cd \de{\bf
v} +
\pder{f_0}{T} \de T, \eenn

\nid where \benn {\bf v}=\int d^d \! p \frac{{\bf p}}{m} f. \eenn

\nid Again, in our model we assume that the electrons have no initial drift
velocity and only
experience a net drift upon application of an electric
field. Rewriting the local distribution as

\benn
f_{\rm loc}=f_0+\pder{f_0}{\ep}\de \mu+ \pder{f_0}{\ep} {\bf p} \cd
\de {\bf v} +\pder{f_0}{\ep}\frac{
(\ep-\mu)}{T} \de T,
\eenn

\nid makes it transparent that it is indeed just the diagonal element
of the corresponding quantum statistical operator in
the limit ${\bf q} \rightarrow 0$. The Boltzmann transport equation,
\eq{boltz}, now
assumes the following form when linearized,

\benn
\pder{\de f}{t}+ \frac{\bf p}{m} \cd \pder{\de f}{\bf r} + e {\bf E} \cd
\pder{f_0}{\bf p}=-\frac{\de
f}{\tau}+\frac{1}{\tau}\pder{f_0}{\ep}\left[ \de \mu + {\bf p} \cd
\de {\bf v} + \frac{(\ep-\mu)}{T}
\de T \right]. \lab{linboltz}
\eenn

As in Sec. \ref{sec:cons} we now examine the consequences of applying the
conservation laws.
The terms $\de \mu, \de {\bf v},$ and $\de T$ are independent of the
electron momentum
and must be determined by the local conservation laws for the collision
invariants : average number, momentum and energy, i.e.

\benn \int {d^d \! p} (f-f_{\rm loc}) \left\{
\begin{array}{cl}
  1 \\
  {\bf p} \\
  p^2/2m
\end{array}\right\} = \ds \left\{
\begin{array}{cl}
  0 \\
  0 \\
  0
\end{array}
\right\}. \eenn

\nid Proceeding in a manner similar to that described in section
\ref{sec:qu}, the solutions are,

\benn
 \de \mu({\bf q},\om)=\frac{\left(\ds B'_0 2 m \mu - B'_2 \right) \left(
\ds \frac{i\om
 C'_2}{\tau q^2 n_0} \de n +UC'_2  \right) + \de n \left(D'_2
 2m \mu - D'_4 \right)}{D'_4 B'_0-D'_2B'_2}, \eenn
 \benn
 \de {\bf v}({\bf q},\om)= \de n\frac{\om}{n_0 q^2}{\bf q}, \eenn and \benn
 \de T({\bf q},\om)= 2mT  \frac{\de n \left(D'_2 + \ds \frac{i \om B'_0
C'_2}{q^2 n_0} \right)+
U C'_2 B'_0}{D'_4B'_0-D'_2B'_2}. \eenn

\nid We have introduced the classical versions of the integral functions
defined in section
\ref{sec:qu}, i.e.

\benn B'_n=\frac{2}{(2 \pi)^d} \int d^dp  |{\bf p}|^n \pder{f_0}{\ep} ,
\eenn \benn C'_n=\frac{2}{(2 \pi)^d} \int d^dp |{\bf p}|^n \pder{f_0}{\bf
p} \cd {\bf q}
\frac{1}{\tilde{\om}-{\bf p}\cd{\bf q}/m}, \eenn and \benn
D'_n=\frac{C'_n+i\om \tau
B'_n}{i \tilde{\om} \tau}. \eenn

\nid Note that the expressions for the thermodynamic variations also
differ from their
quantum counterparts in that terms of order $q^2$ are absent.

We can now obtain the dielectric response of the system with a treatment
similar to that
given above. The DF that follows from solving for $\de n$ in the transport
equation is
identical to \eq{gendie} except that the integral functions, $B_n$, $C_n$
and $D_n$ are
replaced by their classical counterparts, $B'_n$, $C'_n$ and $D'_n$. The
interesting
point here is that functional form is therefore preserved in both the
quantum result and
in the classical approximation. Interestingly, an early attempt to "guess"
a quantum
number-conserving response from the classical response led to erroneous results
\cite{Kliewer}. However the work presented in this paper suggests that
the the correct response may be correctly guessed from the classical
response but only if the classical response can be written in terms of
macroscopic moments of the Lindhard polarizability function.

The semi-classical result follows by taking $f_0$ to be the
equilibrium fermionic distribution. In this case the integral functions,
$B'_n$, $C'_n$ and $D'_n$, and hence the semi-classical DF,
represent the low $q$ expansion of their quantum (unprimed)
counterparts in Sec. \ref{sec:cons} and Sec. \ref{sec:subdie}, respectively.

\section{Hydrodynamics of a degenerate electron gas} \lab{sec:hyd}

A hydrodynamic description is obtained by coarse-graining over the
microscopic degrees
of freedom in such a way that the dynamical equations are expressed in terms of
macroscopic quantities only. The hydrodynamic equations are traditionally
derived from
the classical Boltzmann equation \eq{boltz} and this indeed is the approach
employed by
Tokatly and Pankratov \cite{Tokatly1} who also recently formulated the
hydrodynamic theory of an electron gas to arbitrary frequencies. Here
we provide a derivation from the quantum
description of the microscopic dynamics and begin by writing the
macroscopic quantities
in terms of the one-particle statistical operator,

\benn n({\bf r},t)=\lan {\bf r} | \ha{\rho} | {\bf r} \ran, \eenn

\benn {\bf j}({\bf r},t)=mn({\bf r},t){\bf v}({\bf r},t)=\frac{1}{2} \lan
{\bf r} | \{
\ha{\rho}, \ha{\bf p} \} | {\bf r} \ran, \eenn

\benn U({\bf r},t)=\lan {\bf r} | \ha{U} | {\bf r} \ran, \eenn

\nid and

\benn P_{ij}({\bf r},t)=\frac{1}{2m} \lan {\bf r} | \{ \ha{\rho},
(\ha{p_i}-mv_i)
(\ha{p_i}-mv_i) \} | {\bf r} \ran, \eenn

\nid where $\{, \}$ denotes symmetrization (i.e. anticommutation), $n$
and ${\bf j}$ can be shown to be equivalent to the definitions in \eq{x},
and $P_{ij}$ is the local stress tensor in the comoving frame. Higher
momentum moments
can be neglected in a second order theory of a degenerate electron gas,
i.e. thermal
effects are negligible since $k_BT \ll E_F$. The conserving constraints can
also be
expressed in a similar manner, namely

\benn \lan {\bf r} | \ha{\rho}-\ha{\rho}_{\rm loc} | {\bf r} \ran =0, \eenn

\benn \lan {\bf r} | \{ (\ha{\rho}-\ha{\rho}_{\rm loc}), \ha{\bf p} \} |
{\bf r} \ran=0,
\eenn

\nid and

\benn \lan {\bf r} | \{ (\ha{\rho}-\ha{\rho}_{\rm loc}), \ha{H}_0 \} | {\bf
r} \ran=0.
\eenn

The hydrodynamic equations follow by taking diagonal elements (in position
representation)
of the quantum dynamical equation, which we repeat here,

\benn \pder{\ha{\rho}}{t}+i \left[ \left( \frac{\ha{p}^2}{2m} + \ha{U}
\right), \ha{\rho}
\right]=\frac{\ha{\rho}_{\rm loc}-\ha{\rho}}{\tau}, \lab{mas} \eenn

\nid but symmetrized with the relevant operators from each conservation
law. Before proceeding we note than an alternative approach to
deriving the equations of motion of thermodynamic coordinates from the
quantum Liouville equation, but not necessarily close to equilibrium,
has been discussed via the time-dependent projection operator
technique \cite{Jaynes}.

The first
hydrodynamic equation, the equation of continuity, is simply obtained by
taking the
diagonal element of \eq{mas}, eventually giving

\benn \frac{Dn}{Dt} + n \na_i v_i=0, \lab{fir} \eenn

\nid where $D_t=\partial_t+v_i \na_i$ is the covariant derivative. The
derivation requires use
of the fact that for a one-body potential, $\lan {\bf r} | [
\ha{U},\ha{\rho}] | {\bf r}
\ran=0$ and that $\lan {\bf r}|[\ha{\bf p},\ha{A}]|{\bf r} \ran=i \na
\lan{\bf r}|\ha{A}|{\bf r}\ran$ for any operator $\ha{A}$. The next
hydrodynamic equation follows by anticommuting \eq{mas} with $\ha{\bf p}$
and then
taking the diagonal element, culminating in

\benn mn \frac{D v_i}{D t} + \na_i p + \na_j \pi_{ij} -n \na_i U =0
\lab{sec}, \eenn

\nid where the symmetric stress tensor, $P_{ij}$, has been decomposed into
scalar and
traceless parts, i.e.

\benn P_{ij}=p\de_{ij}+\pi_{ij} \eenn

\nid and

\benn {\rm{Tr}} \pi_{ij}=0. \eenn

\nid The scalar $d$-dimensional pressure of the gas in local equilibrium is
given by

\benn p=\frac{1}{d} {\rm{Tr}} P_{ij} \eenn

\nid which for an ideal and degenerate Fermi gas, of spin degeneracy $2$,
can be calculated to be

\benn p \equiv \frac{k_BT}{V} \ln {\mathcal{Z}}=\frac{2^{2-2/d} \pi
[\Gamma(1+d/2)]^{2/d}}{d+2}
\frac{n_0^{1+2/d}}{m} + O \left( \left (\frac{k_B T}{E_F} \right)^2
\right), \lab{press} \eenn

\nid where $\mathcal{Z}$ is the grand canonical partition function.
The third hydrodynamic equation, obtained by anticommuting \eq{mas} with
$\ha{p_i}\ha{p_j}$, is given to second order by

\benn \frac{DP_{ij}}{Dt}+P_{ij}\na_k v_k + P_{ik} \na_k v_j + P_{kj} \na_k
v_i =
-\frac{\pi_{ij}}{\tau}. \lab{thi} \eenn

\nid By decomposing $P_{ij}$ into $p$, \eq{press}, and $\pi_{ij}$ we
actually get two
equations from \eq{thi}, namely

\benn \pi_{ik} \na_k v_i=0 \lab{fou} \eenn \nid and

\benn \frac{D \pi_{ij}}{Dt} + \pi_{ij} \na_k v_k +
\pi_{ik}\na_{k}v_{j}+\pi_{kj}\na_{k}v_{i}-\frac{2}{d}\de_{ij}\pi_{\mu
k}\na_{k}v_{\mu}+p\left (
\na_{i}v_{j}+\na_{j}v_{i}-\frac{2}{d}\de_{ij}\na_{k}v_{k}\right)=-
\frac{\pi_{ij}}{\tau}.
\lab{anistr} \eenn

The complete set of hydrodynamical equations is then specified by \eq{fir},
\eq{sec},
\eq{press}, \eq{fou} and \eq{anistr}. Traditional Bloch (adiabatic)
hydrodynamics is
recovered by setting $\pi_{ij}=0$. The presence of $\pi_{ij}$ stems from
the anisotropy
of the stress tensor, i.e. the inability of the dynamic electron gas to
achieve local
equilibrium. The content of \eq{anistr} shows that the extent of
anisotropic dynamics
are tempered by relaxation effects which drive the system back to an
isotropic state
of local equilibrium.

We now demonstrate the efficacy of these equations in correcting the
shortcomings of the traditional adiabatic hydrodynamics in predicting the
longitudinal
plasmon dispersion. The simplest route to calculating the dispersion of the
collective
mode, in say the $x$ direction, is to first linearize the hydrodynamic
equations, giving
\benn \pder{\de n}{t}+n_0 \na_x v_x=0, \eenn

\benn mn_0 \pder{v_x}{t}+ \na_x \de p + \na_x \de \pi_{xx}-n_0 \na_x U=0, \eenn

\benn \de p=\frac{(d+2)}{d} \frac{p_0}{n_0} \de n, \eenn

\nid and

\benn \pder{\de \pi_{xx}}{t}+\frac{2d-2}{d}p_0 \na_x v_x = -\frac{\de
\pi_{xx}}{\tau},
\eenn

\nid and then to consider a plane-wave solution $e^{i(q x - \om t)}$ in the
absence of an
external field, i.e. $U=U_{\rm int}$. Ignoring retardation effects, the
potential energy $U({\bf r},t)$ is simply the convolution of the density
perturbation and the Coulomb potential energy, which can be expressed in
Fourier space as $U(q)=-\de n(q) V(q)$, i.e. mean-field Hartree potential.
Solving the linearized
equations then gives the following solution,

\benn \om^2=\om_p^2(q)+\beta^2(\om)q^2, \eenn

\nid where $\om_p(q)$ is given by \eq{wp} and

\benn \beta^2(\om)=\left[\frac{d+2}{d} + \frac{2d-2}{d} \left(
\frac{\om}{\wit} \right) \right]
\frac{p_0}{mn_0}, \lab{beta1} \eenn

\nid or

\benn \beta(\om)=\left[\frac{1}{d}+
\frac{2d-2}{d(d+2)} \left( \frac{\om}{\wit} \right) \right]^{1/2}
v_F. \lab{beta2} \eenn

The coefficient $\beta$ interpolates between the correct limiting forms of
both the
hydrodynamic regime ($\om \tau \ll 1$), in agreement with conventional
adiabatic
hydrodynamic theory, and the collisionless regime ($\om \tau \gg 1$), in
agreement with
RPA \cite{RPAexplain}. In a $d$-dimensional electron gas, the low frequency
limit entails
the fact that collisions are effective in maintaining a local equilibrium
and thus a
conduction electron possesses $d$ translational degrees of freedom. In the
opposite
high-frequency limit the influence of collisions is negligible and the
electron has
effectively one translational degree of freedom, corresponding to
motion in the direction of the applied
field. The notion of effective number of degrees of freedom, $f$, can be made
quantitative by comparing \eq{beta1} with the same result in the adiabatic
limit, i.e.
ignoring the second term in brackets, giving

\benn f(\om)=\frac{d\wit}{\wit +(d-1)\om}. \lab{f} \eenn

\nid The limiting forms of $f(\om)$ and $\beta(\om)$ may be summarized as

\benn
\begin{array}{l}
f(\om)=d, \\
\beta(\om)=(1/d)^{1/2}v_F,
\end{array}
\hspace{1cm} (\om\tau \ll 1) \eenn and
\benn
\begin{array}{l}
f(\om)=1, \\
\beta(\om)=[3/(d+2)]^{1/2}v_F.
\end{array}
\hspace{1cm} (\om\tau \gg 1) \eenn

In three dimensions, \eq{beta2} and \eq{f} agree with those derived by Halevi
\cite{Halevi} who determined the $\om$-dependence of $\beta$ by comparing a
(damped) hydrodynamic DF with the Mermin DF. The need for a frequency-dependent
$\beta$ has also been noted in the context of a hydrodynamic description of
inhomogeneous electron systems \cite{Dobson}. The cross-over between
the collisional and collisionless regime has also recently been
discussed in an interesting paper \cite{elastic}  discussing the viscoelastic
behavior of an electron liquid.

Finally, we note that a popular way of implementing phenomenological
damping in hydrodynamic models of an electron gas through the use of
\cite{Fetter}

\benn \frac{Dn}{Dt} + n \na_i v_i=0, \lab{fird} \eenn

\nid and \benn m \frac{D v_i}{D t} + \frac{\na_i P}{n}  - \na_i U
=-m\frac{v_i}{\tau}
\lab{secd}, \eenn

\nid is simply the hydrodynamic approximation of the Mermin model in the
adiabatic regime
and where all momentum is destroyed during collisions, i.e. \eq{fird} and
\eq{secd} are
obtained from \eq{mas} with the following constraints, namely

\benn \lan {\bf r} | \ha{\rho}-\ha{\rho}_{\rm loc} | {\bf r} \ran =0,
\eenn \benn
\pi_{ij}=0, \eenn \nid and \benn \lan{\bf r} | \{ \ha{\rho}_{\rm loc},
\ha{\bf p} \} | {\bf r}
\ran=0. \eenn

\section{Conclusion}
The relaxation-time approximation has been applied to the quantum dynamics of a
$d$-dimensional electron gas where (average) number, momentum and energy
are conserved during
collisions. It has been shown that the requirement of conservation of all
collision invariants necessitates a state of {\it local} equilibrium, to which
the electrons relax, with a space- and time-dependent chemical potential, drift
velocity and temperature. The ensuing dielectric response has been
determined and
compared with others in the literature, revealing that, in general, that
imposition of the conservation laws tends to make the dielectric response
in the relaxation-time approximation more akin to the RPA dielectric
response than
without them. The FCDF attains the correct static limit and
obeys the perfect screening sum rule.

The resultant plasmon dispersion has been compared
with others and relaxation has been shown to induce softening of the
plasmon with increasing $r_S$ and $1/\tau$. In
both three and two dimensions and in the long wavelength limit, the plasmon
dispersion from FCDF is in
good agreement with that derived from the RPA. In general, correlation
effects are more pronounced in two dimensions than in three and thus further
experimental study
of two-dimensional plasmons ought to provide a significant test of models of
dynamical correlation.

The derivation of the conserving reponse has been repeated but in a
semi-classical
formalism using the Boltzmann transport equation. The functional form
of the ensuing response is then identical to that obtained by quantum
kinetic arguments.

The hydrodynamic equations of a degenerate electron gas are derived from the
quantum microscopic equation of motion using the relaxation time
approximation and
fulfillment of all the conservation laws. The departure of the stress
tensor from a scalar
is due to inertial effects but counteracted by relaxation effects. The
competition
between these two effects gives rise to a generalization of the
hydrodynamics applicable to all frequencies and thus correctly
reproduces the collision-dominated adiabatic limit ($\om\tau \ll 1$)
and the collisionless limit ($\om\tau \gg 1$).

Given that the FCDF aims to incorporate dynamic electron correlation
effects via a phenomenological $\tau$ it is interesting to compare
this with dielectric funtions which include correlation effects via a
many-body field-theoretic approach giving rise to dynamic local field
factors \cite{Singwi}. In principle the two different approaches ought
to exhibit the same qualitative behavior (with respect to differences
from the RPA) and indeed this is what we observe. In this context we
can view the relaxation-time approximation as being a crude way of
including Feynman diagrams beyond the RPA in the proper polarizability
of an electron liquid. To extend the comparisons in a more
quantitative and direct manner further work is required,
i.e. calculation of an expression for $\tau$ and inclusion of exchange
in the FCDF.

Finally, we repeat here that the analysis in this paper explicitly
calculates the response
function only when all the conservation laws are obeyed, as required for a
one-component
system, but, using the results of Appendix \ref{app:lon}, the response
functions can be calculated
which obey fewer and in different combinations. This is the case
required in, for example, two-component systems or when crystalline
effects are incorporated by way of Umklapp scattering.

\section{Acknowledgements}

This work was supported by the National Science Foundation under grant number
DMR-9988576. We thank Prof. B. Tanatar for bringing some recent literature
to our
attention.

\appendix

\section{Thermodynamics of a local equilibrium state} \lab{app:localtherm}

In this appendix we elaborate on what is meant by \textit{local}
values of $\mu$, ${\bf
v}$ and $T$, in the thermodynamic sense.

We first recall how temperature and chemical potential are defined for a large
(macroscopic) system, A, lying in contact with a
heat reservoir, B, with which it can exchange particles and energy.
Thermodynamic
equilibrium of the system+reservior setup is achieved when the entire
entropy of the
system+reservoir is maximized,

\benn \left( \pder{S_{A+B}}{U_A} \right)_{N,V}=\left(
\pder{S_{A+B}}{N_A} \right)_{U,V}=0 \eenn

From the conservation of total particle number, energy and volume we arrive
at the equality of three
thermodynamic variables which we identify as the temperature
($T_A=T_B$) and the chemical
potential ($\mu_A=\mu_B$). Throughout the system $\mu$ and $T$ are
constants independent
of space and time. By appealing to the strong Coulomb attraction
between the electrons and the positive background we can infer that
the velocity
field, ${\bf v}_A$, and hence the pressure, is also constant throughout. The
constant nature of the thermodynamic variables is due to the effect of
relaxation processes which drive the system towards equilibrium. Now, if
the system is
perturbed slightly, these variables may attain a spatial variation and will
still remain
thermodynamically well-defined if the subsection of the system over which
they vary
slightly is much greater than the average distance between
collisions. Thus we require that

\benn |{\bf q}|l_{\text{mf}} \ll 1, \eenn

\nid where ${\bf q}$ is the wavevector of the spatial fluctuations of
$\mu$, ${\bf v}$ and $T$, and $l_{\text{mf}}$ is the mean free path which
can be
approximated as $v_F \tau$. In addition the
time variation of the thermodynamic variables must be slower than the
relaxation time scale in order for the relaxation processes to maintain the
state of
local equilibrium, i.e. \benn \om \tau \ll 1. \eenn \nid This constraint is
consistent
with the fact that the scattering term, and hence the presence of $f_{\rm
loc}$ only
becomes important in the limit that $\tau$ is small.

These considerations suggest that the local equilibrium state may be an
adiabatic one
and indeed we can show that $f_{\rm loc}$ is a small perturbation away from
$f_0$ such
that there is no entropy production. We begin by recalling the Von Neumann
definition of
entropy,

\benn S=-k_B \rm{Tr} (\ha{\rho}_{\rm gc} \ln
\ha{\rho}_{\rm gc}), \eenn

\nid where the trace is carried out over all possible microstates in the
grand canonical
ensemble and $\rho_{\rm gc}$ is the statistical operator of the grand
canonical ensemble. This
expression can be written in terms of the one-particle fermionic
statistical operator, $\ha{\rho}$,
\cite{Thirring}

\benn S=-k_B \rm{tr} [\ha{\rho} \ln \ha{\rho} + (1-\ha{\rho}) \rm ln
(1-\ha{\rho})], \eenn

\nid where $tr$ denotes the trace over one-particle states. In a
non-equilibrium distribution
the occupation and distribution of states acquire both a spatial and time
dependence and
thus, in general, the entropy will be a function of time. The time
variation of $\ha{\rho}_{\rm loc}$ is determined by substitution into
\eq{mas} giving

\benn \pder{\ha{\rho}_{\rm loc}}{t}=-i[\ha{H},\ha{\rho}_{\rm loc}], \eenn

\nid and it follows that when $\ha{\rho}=\ha{\rho}_{\rm loc}$

\benn \pder{S}{t}=0. \eenn

We note that this derivation is essentially the quantum analogue of
Boltzmann's $\Theta$
(or $\mathcal{H}$)-theorem. Thus the variations of $\mu$, ${\bf v}$ and
$T$, that
characterize the local equilibrium state, are constrained to prevent the
production of
entropy in the system. The situation is identical to the propagation of
sound in an
Euler fluid or first sound in superfluid Helium II (to lowest order in
$(\de n/\de
T)_P$): a linearized disturbance away from equilibrium gives rise to an
isentropic
(constant entropy density) longitudinal traveling wave with local
variations in
density, pressure and temperature.

\section{Useful Integrals} \lab{app:int}

The integrals defined in this paper can be expressed as moments of the
polarizability
function, $\chi$, in the Lindhard dielectric constant. The Lindhard
polarizability, in
$d$ dimensions, is given by

\benn \chi({\bf q},\om)=\frac{2}{(2 \pi)^d} \int d^d\! p \, \frac{f_{{\bf p} +
\frac{1}{2}{\bf q}} -f_{{\bf p} - \frac{1}{2}{\bf q}}}{\om - {\bf p} \cd
{\bf q}/m
+i0^{+}}. \eenn

We now define the $n-$th momentum moment of the polarizability function as

\benn \chi_{n}({\bf q},\om)=\frac{2}{(2 \pi)^d} \int d^d\! p \,|{\bf p}|^n
\frac{f_{{\bf
p} + \frac{1}{2}{\bf q}} -f_{{\bf p} - \frac{1}{2}{\bf q}}}{\om - {\bf p}
\cd {\bf q}/m
+i0^{+}}. \eenn

Here we demonstrate how analytical expressions may be obtained in the low
{\bf q} limit
for a degenerate electron gas. In this limit,

\bearraynn {\mathrm{Re}}\chi_{n}({\bf q},\om)&=&\frac{2}{(2 \pi)^d}
{\mathcal{P}} \int
d^d\! p \,
\pder{f}{\ep} \, \frac{|{\bf p}|^n \, {\bf p}\cd{\bf q}}{\om m-{\bf
p}\cd{\bf q}}, \\
{\mathrm{Im}}\chi_({\bf q},\om)&=& \frac{1}{(2 \pi)^{d-1}} \int d^d\!
p \pder{f}{\ep} |{\bf p}|^n {\bf p}\cd{\bf q} \, \de(\om m - {\bf
p}\cd{\bf q}). \eearraynn

\nid In spherical coordinates the real part becomes,

\bearraynn {\mathrm{Re}}\chi_{n}({\bf q},\om)&=&\frac{2m\Omega_{d-1}}{q(2 \pi)^d} \,
{\mathcal{P}} \!
\int dp \,
\pder{f}{\ep} p^{d-2+n} \int_0^{\pi} d\phi \left(\om I[\phi,p]-\frac{p
q}{m} \sin^{d-2} \! \phi \right),\\
\nonumber I[\phi,p]&=&\frac{\sin^{d-2} \! \phi}{\frac{\om m}{p
q}-\cos\phi}, \eearraynn

\nid where $\Omega_d=2 \pi^{d/2}/\Gamma(d/2)$ is the surface area of a unit
d-dimensional
sphere. Upon elementary transformations and exploiting the integral
representation of the
hypergeometric function,

\benn F(a,b,c;z)=\frac{\Gamma(c)}{\Gamma(b)\Gamma(c-b)}\int_0^1 dt \,
t^{b-1}(1-t)^{c-b-1}(1-tz)^{-a}, \hspace{1cm} (c>b>0), \eenn

\nid it is possible to obtain

\benn \int_0^{\pi} d\phi I[\phi,p]=\left\{
\begin{array}{l}
\frd{\sqrt{\pi} \, \Gamma(\frac{d-1}{2})}{\Gamma(d/2)}\left(\pqwm \right)
F\left[1,\frac{1}{2},\frac{d}{2};\left( \pqwm
\right)^{2} \right] \hspace{2cm}(\om m>p\, q) \\
\frd{2\sqrt{\pi} \,
\Gamma(\frac{d-1}{2})}{\Gamma(\frac{d}{2}-1)}\left(\wmpq \right)
F\left[1,2-\frac{d}{2},\frac{3}{2};\left( \wmpq \right)^{2} \right]
\hspace{1cm} (p\,
q>\om m)
\end{array}
\right.
\eenn

\nid and

\benn \int_0^{\pi} d\phi \, \sin^{d-2}\! \phi = \frd{\sqrt{\pi}
\Gamma(\frac{d-1}{2})}{\Gamma(d/2)}. \eenn

If we assume, as is the case for a degenerate electron gas, that $k_BT$ is
much lower
than the Fermi energy then the momentum integral can be simplified with the
substitution
$\pder{f}{\ep}=-\de(\ep-\ep_F)$. The final expressions, at this low
temperature, are then,

\benn \rm{Re} \chi_n({\bf q},\om)= \left\{
\begin{array}{ll}
\frd{m2^{2-d}p_F^{d-2+n}}{\pi^{d/2}\Gamma(d/2)}\left\{1-
F\left[ 1,\frac{1}{2},\frac{d}{2};\left(\pfqwm \right)^2 \right]
\right\} & (\om m > p_F q)\\
\frd{m2^{2-d}p_F^{d-2+n}}{\pi^{d/2}\Gamma(d/2)}\left\{1-\frd{2\Gamma(d/2)}{\Gamma(-1+d/2)}
\left( \wmpfq \right)^2 F\left[1,2-\frac{d}{2},\frac{3}{2};\left(\wmpfq
\right)^2 \right] \right\} & (p_F q > \om m).
\end{array}
\right.
\eenn

\nid The static limit of $\rm{Re} \chi_n$ is thus identified with the
function, $B_n({\bf q})$, for a highly degenerate electron gas.

\section{Generalised solution for thermodynamic fluctuations} \lab{app:lon}

The macroscopic fluctuations of $\de n$, $\de {\bf J}$ and $\de E$, as
defined by \eq{small},
can be obtained by taking the relevant momentum moments of the microscopic
equation of
motion which we rewrite here in terms of ${\bf q}$,

\benn (\om-\De\ep)\dep +\De f U({\bf q})=-\frac{i}{\tau} \left(\dep \!
+ \! \frac{\De f}{\De \ep} \left[\de \mu({\bf q}) + {\bf p} \cd \de {\bf
v} ({\bf q}) +(\frac{p^2}{m} + \frac{q^2}{2m} - \mu) \ds\frac{\de
T({\bf q})}{T} \right] \right).
\lab{aeqmo2} \eenn

However, since we assert that the macroscopic quantities are conserved
before and after
each collision, we must obtain two sets of expressions for each term. The
first set,
determined by the dynamics after each collision process (i.e. right-hand
side of
\eq{aeqmo2}) is given by \eq{small}. The second set follows from solving
the entire
microscopic equation, \eq{aeqmo2}, for $\dep$ and then taking moments, i.e.

\benn \left(
\begin{array}{c}
\de n \\
\de {\bf j} \\
\de E
\end{array} \right)
=-U({\bf q}) \left(
\begin{array}{c}
P[\De \ep] \\
P[{\bf p} \De \ep] \\
P[\frac{p^2}{2m} \De \ep]
\end{array} \right)
-\frac{i}{\tau} \left(
\begin{array}{c}
P[\de \mu({\bf q}) + {\bf p} \cd \de {\bf v}({\bf q}) +(\frac{p^2}{m} +
\frac{q^2}{2m} - \mu) \ds\frac{\de T({\bf
q})}{T}] \\
P[{\bf p}(\de \mu({\bf q}) + {\bf p} \cd \de {\bf v}({\bf q}) +(\frac{p^2}{m}
+ \frac{q^2}{2m} - \mu) \ds\frac{\de
T({\bf q})}{T})] \\
P[\frac{p^2}{2m}(\de \mu({\bf q}) + {\bf p} \cd \de {\bf v}({\bf q})
+(\frac{p^2}{m} + \frac{q^2}{2m} - \mu) \ds\frac{\de T({\bf q})}{T})]
\end{array} \right),
\eenn

\nid where

\benn P[g]=\frac{2}{(2 \pi)^d} \int d^d\! p \, \frac{\De f}{\De \ep} \, \frac
{g}{\wit-\De \ep}. \eenn

\nid Upon expansion of partial fractions we can obtain

\benn \left(
\begin{array}{c}
\de n \\
\de {\bf j}\cd{\bf q} \\
\de E
\end{array} \right)
=-U({\bf q}) \left(
\begin{array}{c}
C_0 \\
\wit C_0 \\
\frac{C_2}{2m}
\end{array} \right)
+\frac{1}{i \wit \tau} \left(
\begin{array}{lll}
B_0+C_0 & C_0\frac{\wit}{q^2} &
\frac{B_2+C_2}{2m}+(\frac{q^2}{8m}-\mu)(B_0+C_0) \\
C_0 \wit & C_0 \frac{{\wit}^2}{q^2} & \frac{C_2}{2m}\wit +
(\frac{q^2}{8m}-\mu)C_0 \wit \\
\frac{B_2+C_2}{2m} & \frac{C_2}{2m} \frac{\wit}{q^2}
&\frac{B_4+C_4}{4m^2}+(\frac{q^2}{8m}-\mu)\frac{B_2+C_2}{2m}
\end{array}
\right) \left(
\begin{array}{c} \de \mu \\
\de{\bf v}\cd{\bf q} \\
\frac{\de T}{T} \end{array} \right), \eenn

\nid where we have, for convenience, taken the component of momentum
density along {\bf q}.
In addition we have assumed that the direction of the momentum density lies
entirely in
the {\bf q} direction, a property that fails for Non-Newtonian
fluids. With this expression, together with \eq{small}, it is possible to
eliminate the conserved
quantities to solve for the thermodynamic variations,

\benn \left(
\begin{array}{ccc}
D_0 & \frac{C_0}{q^2 i \tau} &
\frac{D_2}{2m}+(\frac{q^2}{8m}-\mu)D_0 \\
\frac{C_0}{i \tau} & \frac{\wit C_0}{i \tau q^2}-n_0 &
\frac{C_2}{2m i \tau}+(\frac{q^2}{8m}-\mu)\frac{C_0}{i\tau} \\
\frac{D_2}{2m} & \frac{C_2}{2m q^2 i \tau} &
\frac{D_4}{4m^2}+(\frac{q^2}{8m}-\mu)\frac{D_2}{2m}
\end{array}
\right) \left(
\begin{array}{c} \de \mu \\
\de{\bf v}\cd{\bf q} \\
\frac{\de T}{T} \end{array} \right) =U({\bf q}) \left(
\begin{array}{c}
C_0 \\
\wit C_0 \\
\frac{C_2}{2m}
\end{array}
\right). \eenn

\newpage

\begin{references}
\bibitem{Lindhard} J. Lindhard, Kgl. Danske Videnskab. Selskab, Mat.-Fys.
Medd 28 8 (1954).

\bibitem{Mermin} N. D. Mermin , Phys. Rev. B {\bf 1}, 2362 (1970).

\bibitem{Ash1} P. Garik and N. W. Ashcroft, Phys. Rev. B {\bf 21},
391 (1980).

\bibitem{Ash2} D. M. Wood and N. W. Ashcroft, Phys. Rev. B {\bf 25},
6255 (1982).

\bibitem{Morawetz} A. Selchow and K. Morawetz, Phys. Rev. E {\bf 59},
1015 (1999).

\bibitem{us} G. Atwal and N. W. Ashcroft, Bull. Am. Phys. Soc. {\bf 45}, 867
(2000); {\bf 46} 1147 (2001).

\bibitem{Moranew} K. Morawetz and U. Fuhrmann, Phys. Rev. E {\bf 61}, 2272;
{\bf 62}, 4382 (2000);
K. Morawetz cond-mat/0104229 (unpublished).

\bibitem{Kadanoff} The term ``conserving approximation'' is also used
in the Green's function approach to transport theory, G. Baym and L. P.
Kadanoff,
Phys. Rev. {\bf 124}, 287 (1961). Conservations laws are implemented
there by defining a suitable approximation for the one-particle Green's
function.

\bibitem{Bloch} F. Bloch, Z. Phys. {\bf 81}, 363 (1933).

\bibitem{Tokatly1} I. Tokatly and O. Pankratov, Phys. Rev. B {\bf 60},
15550 (1999), {\bf 62}, 2759 (2000).

\bibitem{Pines} D. Pines and P. Nozieres, \textit{The Theory of
Quantum Liquids}, (W. A. Benjamin, New York, 1996).

\bibitem{Giuliani} G. F. Giuliani and J. J. Quinn, Phys. Rev. B {\bf 29}, 2321 (1984).

\bibitem{Jonson} M. Jonson, J. Phys. C {\bf 9}, 3055 (1976).

\bibitem{Mahan} G. D. Mahan, \textit{Many-particle physics} Plenum
Press, New York, (1990).

\bibitem{Kliewer} K. L. Kliewer and R. Fuchs, Phys. Rev. {\bf 181}, 552 (1969).

\bibitem{Jaynes} E. T. Jaynes, Phys. Rev. {\bf 106}, 620 (1957);
B. Robertson, Phys. Rev. {\bf 144}, 151 (1966); R. Balian, Y. Alhassid,
and H. Reinhardt, Phys. Rep. {\bf 131}, 2 (1986).

\bibitem{RPAexplain} The oft-used statement that RPA is valid only in
the collisionless regime is strictly inaccurate since, of course, the RPA
does incorporate
correlation effects via the mean-field potential (one-body operator) from
Poisson's
equation. By collisionless we thus mean that the two-body operator
term describing scattering process, the right-hand side of \eq{Liou}, is
negligible.

\bibitem{Halevi} P. Halevi, Phys. Rev. B {\bf 51}, 7497 (1995).

\bibitem{Dobson} J.F. Dobson and H.M. Le, Theo Chem {\bf 501}, 327 (2000).

\bibitem{elastic} S. Conti and G. Vignale, Phys. Rev. B {\bf 60}, 7966 (1999).

\bibitem{Fetter} A. L. Fetter, Ann. Phys. (N.Y.) {\bf 81}, 367 (1973).

\bibitem{Singwi} A. Holas, P. K. Aravind, and K. S. Singwi, Phys. Rev. B
{\bf 20}, 4912 (1979); A. Czachor, A. Holas, S. R. Sharma, and
K. S. Singwi,  Phys. Rev. B {\bf 25}, 2144 (1982).

\bibitem{Thirring} W. Thirring, \textit{Quantum Mechanics of Large
Systems}, (Springer-Verlag, 1980).

\end{references}
\end{document}